# Controlling exciton dynamics in two-dimensional $MoS_2$ on hyperbolic metamaterial-based nanophotonic platform

Kwang Jin Lee[1,5], Wei Xin[2,5], Chunhao Fann[1], Xinli Ma[3], Fei Xing[4], Jing Liu[3], Jihua Zhang[1], Mohamed Elkabbash[1], Chunlei Guo[1,2]*

[1] Institute of Optics, University of Rochester, Rochester, New York, USA,
[2] The Guo China-US Photonics Laboratory, Changchun Institute of Optics, Fine Mechanics and Physics, Chinese Academy of Sciences, Changchun, China, 130033
[3] State Key Laboratory of Precision Measurement Technology and Instruments, School of Precision Instruments and Opto-electronics Engineering, Tianjin University, Tianjin, China, 300072
[4] School of Physics and Optoelectronic Engineering, Shandong University of Technology, Zibo, China, 255049
[5] These authors contributed equally to this work

*e-mail: guo@optics.rochester.edu

**The discovery of two-dimensional transition metal dichalcogenides (2D TMDs) has promised next-generation photonics and optoelectronics applications, particularly in the realm of nanophotonics. Arguably, the most crucial fundamental processes in these applications are the exciton migration and charge transfer in 2D TMDs. However, exciton dynamics in 2D TMDs have never been studied on a nanophotonic platform and more importantly, the control of exciton dynamics by means of nanophotonic structures has yet to be explored. Here, for the first time, we demonstrate the control of exciton dynamics in $MoS_2$ monolayers by introducing a hyperbolic metamaterial (HMM) substrate. We reveal the migration mechanisms of various excitons in $MoS_2$ monolayers. Furthermore, we demonstrate the Förster radius of the A-excitons can be increased by introducing HMMs through the nonlocal effects of HMMs due to the Purcell effect. On the other hand, the diffusion coefficient is unchanged for the C-excitons on HMMs. This study provides a revolutionary step forward in enabling 2D TMD nanophotonics hybrid devices.**

With the explosive research activities since the discovery of graphene, two-dimensional (2D) materials have emerged as one of the most exciting areas studied in science and engineering[1–5]. Among them, 2D transition metal dichalcogenides (TMDs) have attracted a great amount of attentions and been considered as an ideal material for nanophotonic and optoelectronic applications due to their remarkable optical and electronic properties, such as, higher photoluminescence efficiency due to direct bandgap and existence of light-valley interactions[6-11]. Atomically-thin monolayer TMDs have strongly bounded excitons because of the enhancement in quantum confinement and Coulomb interactions, and this strong bonding dominates most optical and electronic effects. In general, exciton binding energy in TMD monolayers is an order of magnitude higher than that of previously investigated 2D quantum well structures, which leads to their unique optoelectronic characteristics and makes TMDs an ideal platform for exploring exciton dynamics (ED) that is essential for photo-current conversion processes and novel optoelectronic applications[12,13]. An analogue can be seen in organic semiconductors, which also have large exciton binding energies due to their low dielectric constants and this effect incites a large amount of exciton dynamics studies in organic photovoltaic operation[14-17]. Therefore, a thorough characterization of ED is of paramount importance for improving light-harvesting applications as well as revealing fundamental mechanism of carrier dynamics in 2D TMD materials. Furthermore, understanding and controlling ED in these materials when integrated into a nanophotonic platform has never been explored before.

Engineering light-matter interactions has been realized using nanophotonic structures, e.g., metamaterials and engineered materials with tailored optical properties[18-20]. Particularly, metamaterials have been used in optoelectronic devices[21,22], optical sensing[23], plasmonic lasers[24] and Raman spectroscopy[25]. Among various types of metamaterials, hyperbolic metamaterials (HMMs) have been extensively studied over the past few years due to their unusual optical properties from the high-*k* states[26-30]. HMM structures that have been shown to exert nonlocal effects on the photophysical properties of their surrounding environment have recently been reported[31,32], which suggests that TMDs can be drastically altered without modifying the material itself, but instead by incorporating a TMD on a HMM.

$MoS_2$ monolayers exhibit two typical band-edge excitons, A- and B-excitons, resulting from transitions between the conduction band minimum and spin-orbit split valence band maximum near the *K* point. In addition, recent studies observed another exciton, labelled as C-excitons, with a strong and broadband absorption at higher energies. C-exciton states are attributed to the band nesting effect, i.e., transition arising from the maxima in the joint density of state (JDOS) when the conduction and valence bands are parallel in a region between *K* and $\Gamma$ points[33-37]. Unlike A- and B-excitons, C-excitons have no photoluminescence. Although a few studies have attempted to address some aspects of ED in TMD monolayers[38-40], the underlying mechanism still remains unclear. Particularly, the ED in TMD monolayers has never been studied on a nanophotonic platform.

In this article, we study comprehensively the nonlocal effect of HMMs on ED in $MoS_2$ monolayers. We show that ED in the A and C-excitons have very different dynamics; the migration of A- excitons is mainly through a single-step Förster-type resonance energy transfer (FRET) whereas multi-step diffusion process is responsible for C-excitons. We also find that the Förster radius increases in the presence of the HMM substrates in the hyperbolic dispersion region, but the diffusion constant is not affected by the HMMs. We elucidate that the increased Förster radius comes from the nonlocal effects of HMMs from the Purcell effect. We note there has been ongoing debates in understanding FRET in complex photonic environment[41], and this study provides conclusive evidence to address these issues.

**Sample configuration and transient absorption measurement**

$MoS_2$ monolayer was prepared on silicon substrates by means of chemical vapor deposition. Single-layer samples were identified by optical microscopy and Raman spectra shown in SI-figure 1. Multi-layered HMMs consisting of 5 pairs of alternative Ag-$TiO_2$ layers with different fill factors ($f$=0.2, 0.5 and 0.8) were fabricated by electron beam evaporation. Detail sample configurations are described in SI-figure 2. We confirmed that the peaks of Raman spectra were not altered with HMM substrates (SI-Figure 3). In our design, a 10-nm thick $Al_2O_3$ layer was deposited on top of the stack to avoid the convolution of other processes such as charge transport between $MoS_2$ and HMMs. (Band alignments are illustrated in SI-figure 4). Figure 1a schematically displays the sample configuration for $MoS_2$ monolayer deposited on a HMM structure with $f$ = 0.5 (10 nm thickness of each layer). To observe the ED, we used exciton-exciton annihilation (EEA) method by performing ultrafast transient absorption (TA) experiment based on the pump-probe technique described below. Figure 1b shows the absorption and photoluminescence spectra of $MoS_2$ monolayer. The two absorption peaks at 1.87 eV and 2.05 eV correspond to A- and B-excitons of $MoS_2$ monolayers, respectively. The broad absorption band above 2.8 eV corresponds to the non-emissive C-excitons. The photoluminescence peak and shoulder at 1.84 eV and 2.01 eV correspond to A- and B-excitons, respectively. Figure 1c presents the real part of an effective dielectric constant of HMMs along the transverse direction calculated by effective medium theory. HMM with $f$ = 0.8 ($f$ = 0.2) shows hyperbolic (elliptic) dispersions region for both A-and C-excitons, whereas HMM with $f$ = 0.5 exhibits hyperbolic (elliptic) dispersion for A- (C-) excitons.

**Exciton dynamics for A- and C-excitons**

Ultrafast TA experiments were carried out to analyze the ED of $MoS_2$ monolayers by measuring relative reflection ($\Delta R/R$). The pump beam at 2.25 eV (3.05 eV) and probe beam at 1.85 eV (3.05 eV) were chosen for A- (C-) excitons. The pump fluence for A- and C-excitons were adjusted to obtain the same initial exciton densities ($n_0$) immediately after the excitation by the pump. (SI-figure 5) Figures 2a and 2b show the normalized TA kinetics of A- and C-excitons in $MoS_2$ monolayer on Si substrate without metamaterials for different exciton densities. At the lowest initial exciton density ($n_0$=0.06 × $10^{12}$ $cm^{-2}$), TA kinetics for both A- and C-excitons are fitted by a mono-exponential decay functions with characteristic time (τ) of about 186 ps and 213 ps, corresponding to the intrinsic exciton lifetimes. C-excitons have a relatively longer lifetime than A-excitons, and this is consistent with previous works, which show that favorable band alignment and transient excited state Coulomb environment could lead to a longer lifetime of C-excitions[33,36]. As $n_0$ increases, the decay of A-excitons deviates from a mono-exponential fitting due to an EEA taking place where two excitons are sufficiently close to interact and to generate a single exciton with a higher energy. Using bi-exponential decay fitting, we found that the short time constant ($\tau_1$) decreases with $n_0$. On the other hand, the longer time constant ($\tau_2$) is almost independent of $n_0$, indicating that $\tau_1$ represents EEA phenomenon and $\tau_2$ corresponds to the intrinsic exciton lifetime (SI-figure 6). For C-exciton, we observed a relatively weak dependence on $n_0$, which is also consistent with previous work suggesting that the exciton dissociation occurs efficiently, in agreement with the self-separation of photocarriers in the nesting region in the momentum space[37]. In addition, for A-excitons, we note that figure 2a is consistent with previous study.[38]

Figures 2c and 2d display the TA decays for A- and C-excitons in the initial time range (up to ~ 100 ps). To analyze the EEA behaviour, we consider the rate equation of EEA described by[42,43]

$$\frac{d}{dt}n(t) = -\frac{n(t)}{\tau} - \frac{1}{2}\gamma(t)n(t)^2 \tag{1}$$

where $n(t)$ is the exciton density at a delay time $t$ after the excitation, $\gamma(t)$ is the annihilation rate coefficient and τ is the exciton lifetime at the low exciton density limit ($\tau_2$). The factor 1/2 represents that only one exciton is left after EEA. We note that EEA is dominant over the Auger recombination in this structure[35]. In general, EEA process can be classified by three different mechanisms: three-dimensional (3D) and one-dimensional (1D) multi-step exciton diffusions and a single-step FRET[42]. The exciton diffusion model assumes that the excitons move in random walk in many steps towards each other before the annihilation takes place. On the other hand, FRET model considers that annihilation occurs directly via long-range energy transfer processes (SI-Figure 7). FRET strongly depends on the overlap between the emission spectrum of the donor and the absorption spectrum of the acceptor.

For 2D $MoS_2$ monolayers, we only need to consider FRET and 1D exciton diffusion mechanisms. $\gamma(t)$ is given by $\alpha \cdot t^{-1/2}$ where $\alpha = R_F^2 \pi^{3/2}/2\tau^{1/2}$ for the FRET model with $R_F$ is the Förster radius and $\alpha = (8D/\pi)^{1/2}/aN_0$ for the 1D diffusion model with the diffusion constant $D$, lattice constant $a$ and molecular density $N_0$. From these relations, Eq. (1) can be solved as[42,43],

$$n(t) = \frac{n_0 e^{-t/\tau}}{1 + \beta\, \mathrm{erf}\left(\sqrt{\frac{t}{\tau}}\right)} \tag{2}$$

where 'erf' is the error function. The coefficient $\beta$ is expressed by $n_0 R_F^2 \pi^2/4$ and $n_0 l_D/aN_0$ for FRET and the 1D diffusion process, respectively. $l_D$ is the diffusion length defined as $(2D\tau)^{1/2}$. The $a$ and $N_0$ of $MoS_2$ monolayers were taken as 3.16 Å and 5.7 × $10^{14}$ $cm^{-2}$, respectively. Here, it is worth noting that $n(t)$ for both the FRET and 1D exciton diffusion models have the same mathematical structure.

The solid curves in figures 2b and 2e represent the fits based on Eq. (2). For C-excitons, FRET was excluded due to their non-emissive property.[42] The diffusion constant $D$ determined from the fits of the TA decays based on the 1D diffusion model is plotted in figure 2f. Here, the exciton lifetime τ without annihilation was kept as a constant ($\tau$ = 213 ps) and thus was not a fitting parameter. For the highest value of $D$, we can also estimate $l_D$ as ~88 nm, which is considerably longer than that for organic semi-conductors and polymers as shown in SI-Figure 8. To better understand the diffusion mechanism, we plot $D$ versus the inverse of square root of the pump fluence that corresponds to the inverse square root of temperature ($T^{-1/2}$) in SI-Figure 9. The linear behaviour of $D$ with $T^{-1/2}$ indicates that the diffusion process takes place through phonon scattering[43,44].

Due to the strong overlap between the absorption and emission spectra of A-excitons, annihilation via FRET is likely to dominate the EEA process. Figure 2e shows $R_F$ as a function of $n_0$ for A-exciton. We find that $R_F$ value is around 6.0 ~ 6.4 nm and hardly depends on $n_0$, which is consistent with the fact that exciton density is not related to $R_F$. Since FRET is not a prerequisite for A-exciton emission, we can consider the diffusion process as the exclusive mechanism for A-excitons. However, when we plot $D$ instead of $R_F$ as shown in SI-Figure 10, we noticed $D$ does not depend on $n_0$, indicating that diffusion is not the main mechanism for A-excitons.

**Effect of hyperbolic metamaterials**

Figure 3 shows the behavior of time constants with different substrates. We note that $\tau_1$ and $\tau_2$ remain constant for all substrates with no overlapping hyperbolic dispersion, while a discernible decrease in $\tau_1$ and $\tau_2$ is observed for HMM with $f$=0.5 and 0.8. of the decrease of $\tau_2$ from 186 ps to 150 ps can be easily understood in terms of the Purcell factor (PF) enhancement based on the high local density of optical states provided by HMMs. Here, we obtain PF of ~1.24 from basic relationship given by $\tau_2^{Si}/\tau_2^{HMM}$. (SI-Table 1) Interestingly, a shortening of $\tau_1$ due to the hyperbolic dispersion indicates that the nonlocal effect of HMM based on the PF enhancement clearly affect ED occurring through FRET. The 1$p$ substrate consists of a single pair of 10 nm thick Ag/$TiO_2$ films with a 10-nm $Al_2O_3$ serves as a control sample showing the relatively unmodified decay kinetics of $MoS_2$. For C-excitons, while $\tau_1$ appears to be independent of the substrates, we observed an increase in $\tau_2$ within experimental error. The increase in $\tau_2$ is somewhat similar to the increase in the charge recombination time with the HMM substrates observed in previous studies[31]. The entire TA data were plotted in SI-Figure 11. In SI-Figure 12, we adjusted pump fluence to obtain the same $n_0$ by considering field intensity variation in the presence of HMM structures so that we could exclude additional effects that can affect $n_0$.

In figures 4a and 4b, we plot $R_F$ and $D$ as functions of $n_0$ for Si and HMM with $f$=0.2 and 0.8 substrates, respectively. We can see that HMM with $f$=0.5 is almost identical to $f$=0.8. Figure 4a exhibits an enhancement in $R_F$ for the A-excitons in the HMM hyperbolic dispersion regimes. We can explain this interesting result in terms of the nonlocal effect of HMMs based on the PF enhancement, which was shown previously that the nonlocal effect of HMMs could lead to a decrease in the refractive index of the environment[32]. Here, we equivalent the problem as the emitter was placed in a homogenous medium with modified $n$. Based on this discussion, we showed that PF is inversely proportional to $n^3$ (Method). We also apply this concept to FRET, and we obtained the relationship between $R_F$ and PF (denoted as $F_p$) as follows, (Method section).

$$R_F \propto F_p^{\frac{2}{9}} \tag{3}$$

This relation presents a quantitative enhancement factor $R_F$ by 1.05, which is displayed as the open circles in figure 4a. Surprisingly, the predicted values based on the nonlocal effect of HMMs are almost consistent with the experimental values. We note that the current $MoS_2$-HMM hybrid systems are an ideal platform to investigate the fundamental relationship between FRET and photonic environment by excluding quenching effects such as donor-HMM coupling and the charge transport between $MoS_2$ and HMMs. In case of the diffusion processes, as shown in figure 4b, there is no noticeable change in the presence of HMMs, which can be explained by the fact that diffusion processes are not relevant for light-matter interactions. Figure 4c illustrates the overall ED for both A- and C-excitons in $MoS_2$ monolayers.

## Conclusion and Discussion

In conclusion, we reveal the different underlying mechanisms for ED in $MoS_2$ monolayers for the first time such that a single-step Förster-type resonance energy transfer is predominant for A-excitons while a multi-step diffusive motion is responsible for C-excitons. Furthermore, we provide TMDs with a range of nanophotonic platforms using HMMs with different fill factors and find an increase in the Förster radius for A-excitons when A-exciton spectral region lies in the hyperbolic dispersion region. The consistency between experimental results and theoretical rationalization in terms of the nonlocal effect of HMMs based on the Purcell effect shows a decisive evidence that HMMs can alter the FRET efficiency. Our study not only identifies the exact physical mechanisms for ED in $MoS_2$ monolayers, but also provides a conceptual basis for how FRET is affected by HMM structures. Fundamentally, ED and FRET play vital roles in a broad range of technological fields, such as material science, nanophotonics, biology, and optoelectronics. Technologically, our work demonstrates a novel way to nano-engineering 2D materials with a metamaterial-based nanophotonic platform, which will advance the applications of 2D materials in photonics, optoelectronics, and meta-devices.

## Method

### Sample preparation

The preparation of a single-layer $MoS_2$ on silicon substrate (<100>, ≈300 nm $SiO_2$) is based on traditional chemical vapor deposition (CVD) method (high temperature, Argon environment for 2.5 h). $MoO_3$ (99.99%, Aladdin) and S (99.99%, Alfa Aesar) powders were chosen as precursor materials. The PTAS (perylene-3, 4, 9, 10-tetracarboxylic acid tetra potassium salt) was also dropped on the substrate as the seeding promoter to increase the nucleation. After that, by using the wet transfer method, the CVD single-layer $MoS_2$ was then transferred onto the target hyperbolic metamaterials (HMM) that consists of $Al_2O_3$ film, multi-layered Ag-$TiO_2$ layers, and silica substrate. To remove the surface contaminants and make a close contact between $MoS_2$ and the substrate, annealing at 300 °C for 1 h was used.

### Absorption, Photoluminescence and Raman spectra measurement

The UV−vis absorption of CVD single-layer $MoS_2$ on silica and HMM substrates were both carried out using a Cary 5000 UV−visible −NIR spectrometer (Agilent). The Raman and photoluminescence signals were collected by a LabRAM HR Evolution Raman spectrometer (Horiba Jobin Yvon, 100× objective (N.A. = 0.9) and 1800 lines $mm^{-1}$ gratings). The excitation laser was 532 nm at a power of 0.1 mW.

### Transient absorption measurement

Ultrafast transient absorption measurements were carried out using a femtosecond pump-probe setup. ATi:sapphire regenerative amplifier system is used that delivers 2 μJ pulses at 67-fs in duration, with a wavelength centered at 800 nm (1.55 eV) and a repetition rate of 1 kHz. The laser output is split into a pump and a probe beam by a beam splitter. For the A-exciton measurements, the pump and probe beams were obtained by each passing through a 5-mm thick sapphire window to generate white-light-continuum, followed by further passing through a band-pass filter centered at 2.25 eV for the pump and 1. 85 eV for the probe. Each bandpass filter has a bandwidth of 0.03 eV. For the C-exciton measurements, second harmonic (3.05 eV) is generated from the fundamental laser beam using a BBO crystal and the beam is further spilt into a pump and a probe. The Pump beam was modulated using a mechanical chopper at 220 Hz and the relative reflection $\Delta R/R$ of the probe beam as a function of the delay time was further read out with a photodiode and a lock-in amplifier. The relative reflectance is given by $\Delta R/R = (R_{on} - R_{off})/R_{off}$, where $R_{on}$ and $R_{off}$ are the sample reflectance with the pump beam on and off, respectively. The beam sizes of the pump and probe beams are 100 μm and 80 μm in diameter, respectively. The pump power was varied using variable neutral density filters.

**Calculation of exciton density**

The initial excitation density ($n_0$) immediately after the pump excitation can be calculated by the following equation[43],

$$n_0 = \frac{f(1-10^{-A})}{E_{pump}}$$

(M1)

where $f$ is the pump fluence, $E_{\text{pump}}$ is the photon energy of pump beam and $A$ is the absorptance for a given energy given by absorption coefficient times layer thickness.

**Derivation of $\gamma_F(t)$ for two-dimensional case**

To derive a rate equation for a single-step annihilation mechanism through FRET, we started with the trapping problem based on the Förster model where a donor molecule is surrounded by acceptor molecules with the volume concentration $n_T$. In this case, the final decay rate of a single donor molecule can be represented by a summation of the intrinsic decay rate and the energy transfer rates between a single donor molecule to all the surrounding acceptor molecules. Hence, we can obtain the time dependent exciton density on the donor molecules by averaging over all distributions of donor molecules as following[42]

$$n(t) = n_0 \cdot \exp\left(-\frac{t}{\tau} - \alpha \cdot n_T \sqrt{\frac{\pi t}{\tau}}\right)$$

(M2)

Depending on the dimensionality, coefficient α is given by

$$\alpha_{3D} = \frac{4}{3}\pi R_F^3, \quad \alpha_{2D} = \pi R_F^2$$

(M3)

$\alpha_{3D}$ and $\alpha_{2D}$ are the total volume and area at which FRET can take place for three- and two-dimension, respectively. In this work, we adopt $\alpha_{2D}$ considering the current system we are dealing with is a two-dimensional monolayer.

Derivation of $n(t)$ with respect to $t$ yields a rate equation as following

$$\frac{d}{dt}n(t) = -\frac{n(t)}{\tau} - \gamma_F(t) n_T n(t)$$

(M4)

where the exciton annihilation rate $\gamma(t)$ is defined as

$$\gamma_F(t) = \frac{1}{2}\pi R_F^2 \sqrt{\frac{\pi}{\tau t}}$$

(M5)

**Förster radius**

Förster radius is generally expressed as

$$R_F^6 = \frac{9\eta_D \kappa^2}{128\pi^5 n^4}\int d\lambda \lambda^4 F_D(\lambda)\sigma_A(\lambda)$$
(M6)

where $\kappa$ is the dipole orientation factor, $n$ is the refraction index of the environment, $\lambda$ is the wavelength, $F_D$ is the normalized emission spectrum, and $\sigma_A$ is the absorption cross-section. $\eta_D$ is the quantum yield expressed as the ratio of the rate of radiative recombination to the total rate of exciton decay. We note that $\eta_D$, $F_D$ and $\sigma_A$ are the individual properties for isolated configuration so that the above equation is not associated with the density of donor and acceptor.

**Relationship between Purcell factor and Förster radius**

When an emitter is located at a distance of $d$ away from a metallic film with a perpendicular or parallel orientation to the interfaces, the Purcell factors can be written as[28],

$$F_{\mathrm{perp}} = 1-\eta+\frac{3}{2}\eta\,\mathrm{Re}\int_0^\infty \mathrm{d}k_x \frac{1}{k_z}\left(\frac{k_x}{k_0}\right)^3 \frac{1}{n^3}\left[1+r_\mathrm{p} e^{2ik_z d}\right]$$
$$F_{\mathrm{para}} = 1-\eta+\frac{3}{4}\eta\,\mathrm{Re}\int_0^\infty \mathrm{d}k_x \frac{1}{n}\frac{k_x}{k_0 k_z}\left(1+r_\mathrm{s} e^{2ik_z d}+\frac{k_x}{n^2 k_0^2}\left[1-r_\mathrm{p} e^{2ik_z d}\right]\right)$$
(M7)

where $\eta$ is the internal quantum efficiency of the emitters in a free space; $k_x$ ($k_z$) is the component of the wavevector along the $x$ ($z$) axis, and $k_0$ is the magnitude of the wavevector in vacuum; $r_p$ and $r_s$ is the reflection coefficient at the interfaces for p- and s-polarized waves, respectively. In the presence of HMMs, both $k$ and $r$ should be altered so that Purcell factor can be enhanced. In this study, we assume that the emitter is located in a homogeneous medium with a nonlocally modified refractive index, taking into account the overall alteration of $k$ and $r$ by the HMMs. Therefore, we can write the Purcell factor as follows:

$$F_{\mathrm{perp}} = \frac{3}{2}\mathrm{Re}\int_0^\infty \mathrm{d}k_x \frac{1}{k_z}\left(\frac{k_x}{k_0}\right)^3 \frac{1}{n_{\mathrm{HMM}}^3}$$
$$F_{\mathrm{para}} = \frac{3}{4}\mathrm{Re}\int_0^\infty \mathrm{d}k_x \frac{1}{n_{\mathrm{HMM}}}\frac{k_x}{k_0 k_z}\left(1+\frac{k_z^2}{n_{\mathrm{HMM}}^2 k_0^2}\right)$$
(M8)

where $\eta$ is assumed to be unity and $n_{\mathrm{HMM}}$ is the nonlocal refractive index by taking into account the impact of the HMM substrate on $k_0$, $k_x$, and $k_z$. Also, $r_p$ and $r_s$ are zero because we assume that the emitter is placed in a uniform media with refractive index $n_{\mathrm{HMM}}$. Hence, we can obtain the following relation between the Purcell factor and $n_{\mathrm{HMM}}$,

$$F_\mathrm{p}^{\mathrm{HMM}} \propto n_{\mathrm{HMM}}^{-3}$$
(M9)

This relationship shows that the Purcell factor enhancement leads to a decrease in nonlocal refractive index. The above equation can be rewritten as

$$n_{\mathrm{HMM}} \propto \left(F_\mathrm{p}^{\mathrm{HMM}}\right)^{-\frac{1}{3}}$$
(M10)

We can reconsider Eq.(M6) to extract the relationship between nonlocal refractive index and the Förster radius. In the presence of HMMs, we can also consider the physically equivalent homogeneous configuration with nonlocally modified refractive index $n_{\mathrm{HMM}}$

$$R_{\mathrm{F}}^{\mathrm{HMM}} = \alpha\left(n_{\mathrm{HMM}}\right)^{-\frac{2}{3}}$$
(M11)

where

$$\alpha = \left(\frac{9\eta_{\mathrm{D}}\kappa^2}{128\pi^5}\int d\lambda\lambda^4 F_{\mathrm{D}}(\lambda)\sigma_{\mathrm{A}}(\lambda)\right)^{\frac{1}{6}}$$
(M12)

Based on Eq. (M10) and (M11), we can obtain the final relationship between nonlocal refractive index and the Förster radius as following

$$R_{\mathrm{F}}^{\mathrm{HMM}} \propto \left(F_p^{\mathrm{HMM}}\right)^{\frac{2}{9}}$$
(M13)

**Acknowledgments**
This research was supported by Bill & Melinda Gates Foundation, National Science Foundation, and National Natural Science Foundation (NSFC, Grant Nos. 11804334).

## Additional Information

### Author contributions

K.J.L. and C.G conceived and designed the experiments. K.J.L. carried out the transient absorption measurements and analyzed the entire data and wrote the manuscript. X.M., F. X., and J. L. prepared $MoS_2$ monolayers. W.X. prepared $MoS_2$-HMM hybrid structures and characterized the entire samples. C.H., J.Z., and M.E. prepared and characterized metamaterials samples. C.G. supervised whole project. All authors commented the manuscript.

### Competing Financial Interests

The authors declare no competing financial interests.

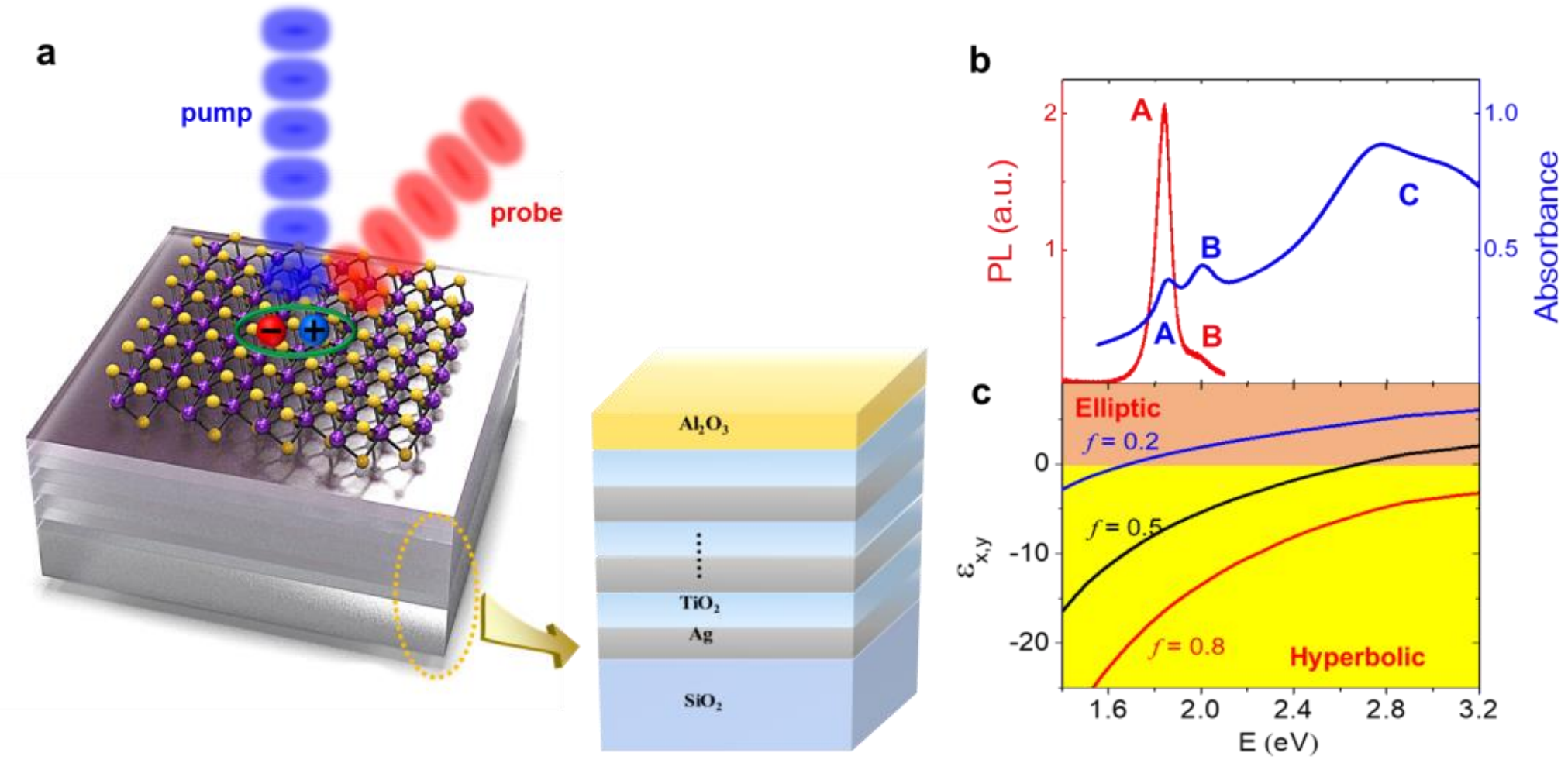

**Figure 1. Sample configuration and characterization. a,** Schematics of the experimental configuration for a $MoS_2$ monolayer with exciton dynamics based on the transient absorption measurements. The $MoS_2$ monolayer is deposited on a multilayered HMM structure that consists of 5 pairs of Ag-$TiO_2$ layers covered by a $Al_2O_3$ film to block any charge transport **b,** Absorption and photoluminescence spectra of $MoS_2$ monolayers with A-,B- and C-excitons. **c,** the Real part of the transverse effective dielectric function of HMM for three different fill factors (*f*=0.2, 0.5, 0.8)

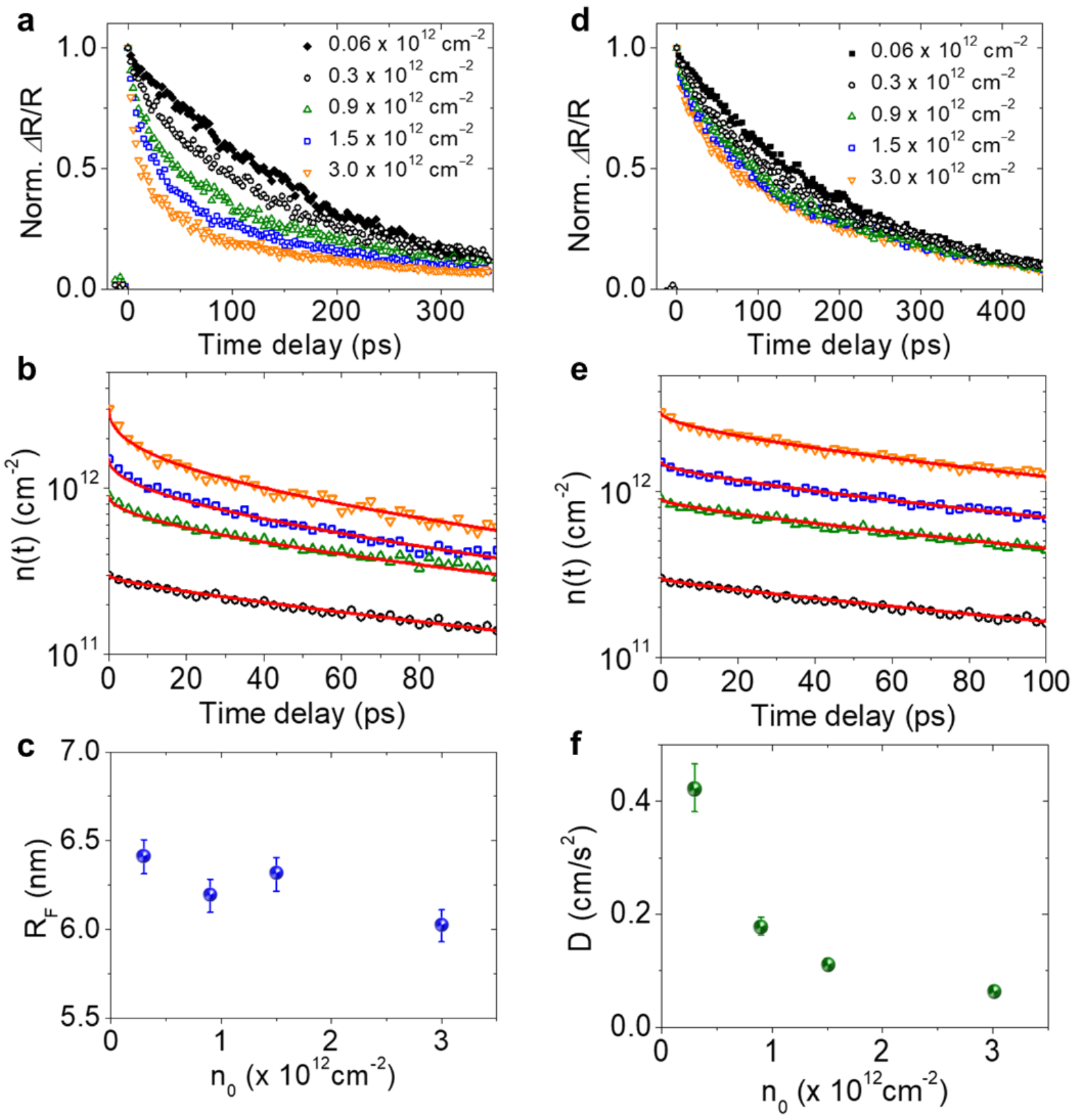

**Figure 2. Transient absorption decays and fitting curves based on exciton-exciton annihilation. a~c**: A-excitons, **d~f**: C-excitons. a,d, Normalized transient absorption decay of A- and C-excitons respectively for several initial exciton densities. b,e, Exciton decays for A- and C-excitons respectively in the initial time range (up to 100 ps) with fitting curves based on Eq. (2). c, f, the Forster radii and the diffusion constants for A- and C-excitons with initial exciton density $n_0$.

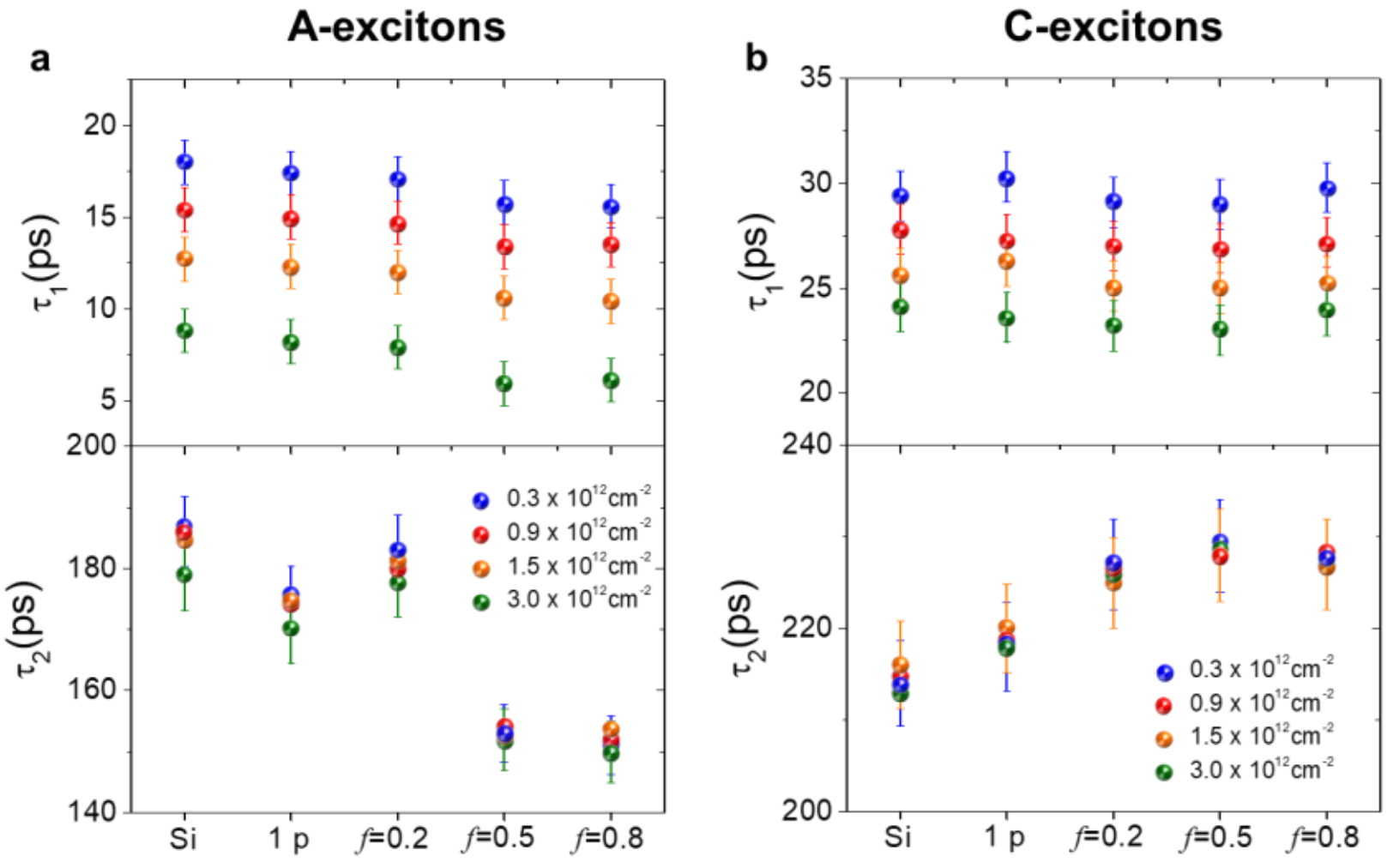

**Figure 3. Behaviors of time constants for A- and C-excitons with different substrates a,** Short ($\tau_1$) and long ($\tau_2$) characteristic time constants of A-exctions with different substrates for several initial exciton densities. **b,** Short ($\tau_1$) and long ($\tau_2$) characteristic time constants of C-exctions with different substrates for several initial exciton densities.

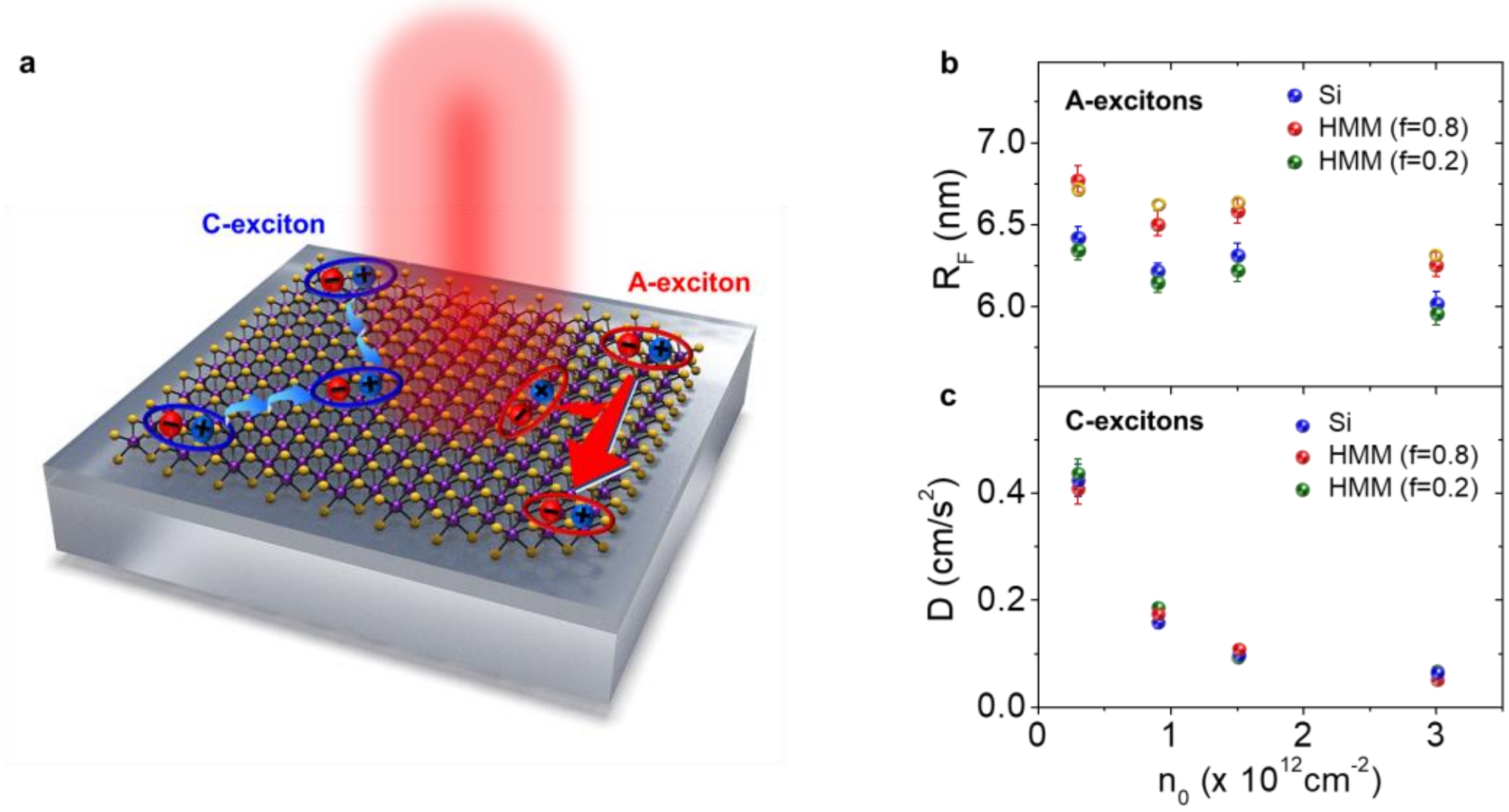

**Figure 4. Underlying mechanism for exciton dynamics and behaviors of the Förster radius and diffusion constant a,** Shematics shows the different migration mechanisms for A- and C-excitons **b,** Förster radius and **c,** the Diffusion constants for A- and C-excitons as a function of the initial exciton density on different substrates (Si, HMM with *f*=0.2 and 0.8). The overall measured behaviors for *f*=0.5 are almost identical to those for *f*=0.8.